\documentclass[journal,article,submit,pdftex,moreauthors]{Definitions/mdpi} 

\firstpage{1} 
\pubvolume{1}
\issuenum{1}
\articlenumber{0}
\pubyear{2023}
\copyrightyear{2023}
\datereceived{ } 
\daterevised{ } 
\dateaccepted{ } 
\datepublished{ } 
\hreflink{https://doi.org/} 

\newcommand{\be}{\begin{equation}}
\newcommand{\ee}{\end{equation}}
\newcommand{\ba}{\begin{eqnarray}}
\newcommand{\ea}{\end{eqnarray}}

\newcommand{\la}{\label}

\def\x{\xi}

\Title{The complex heavy quark potential in an anisotropic quark-gluon  plasma$^\dagger$}

\TitleCitation{The complex heavy quark potential in an anisotropic quark-gluon  plasma}

\Author{Ajaharul Islam $^{1}$*\orcidA{}}

\AuthorNames{Ajaharul Islam}

\AuthorCitation{Islam, A.}

\address{%
$^{1}$ \quad Department of Physics, Kent State University, Kent, OH 44242, United States\\}

\corres{Correspondence: aislam2@kent.edu}

\firstnote{Presented at Hot Quarks 2022—the 9th edition of the Workshop for Young Scientists on the Physics of Ultra-relativistic Nucleus-Nucleus Collisions, Dao House, Estes Park, Colorado, USA, 11–17 October 2022.} 

\preto{\abstractkeywords}{\nolinenumbers}
\abstract{We present a technique to represent anisotropic heavy-quark potentials as effective potentials. This involves employing an effective screening mass linked to the quantum numbers $l$ and $m$ of a specific state. Our approach utilizes the resulting 1D effective potential model, enabling the solution of a 1D Schrödinger equation. Remarkably, this model accurately reproduces the energies and binding energies of low-lying heavy-quarkonium bound states in 3D, including the differentiation of various p-wave polarizations. The derived 1D effective model offers a means to incorporate momentum anisotropy effects into simulations of heavy-quarkonium dynamics in the quark-gluon plasma within open quantum systems.}

\keyword{Quark-gluon plasma, Heavy quarkonium, complex potential, anisotropic plasma} 

\begin{document}
\section{Introduction}
The survival probability of the heavy quarkonium states has been widely used as a sensitive probe to study the quark-gluon plasma (QGP) formed in relativistic heavy-ion experiments at RHIC and LHC~\cite{Matsui:1986dk, Karsch:1987pv}. Due to the non-relativistic characteristics of heavy quarkonium states, it is possible to determine their in-medium properties, including masses and decay rates, through the solution of a Schrödinger equation incorporating a complex heavy-quark (HQ) potential. The real component of the HQ potential yields the binding energy, while the imaginary component offers insights into the decay of a quarkonium state through wave function decoherence ~\cite{Laine:2006ns,Brambilla:2008cx, Escobedo:2008sy,Brambilla:2013dpa}. The HQ potential at short distances can be derived by applying hard-thermal-loop (HTL) resummed perturbation theory in the weak coupling limit. To make a phenomenological study of the effect of momentum-space anisotropies in heavy-ion collisions, we consider the following spheroidal distribution function ansatz in the local rest frame (LRF) of the QGP~\cite{Romatschke:2003ms}
\be\label{anisodis}
f_{\rm aniso}^{\rm LRF} ({\bf k})\equiv f_{\rm iso}\!\left(\frac{1}{\lambda}\sqrt{{\bf k}^2+\xi ({\bf k}\cdot {\bf n})^2}\right)\,.
\ee

This form takes into account the rapid longitudinal expansion of the QGP  at early times and allows for explicit pressure anisotropies in the LRF~\cite{Strickland:2014pga}. Here, $f_{\rm iso}$ is an arbitrary isotropic distribution function, $\lambda$ is a temperature-like scale, which becomes the temperature $T$ of the system in the thermal equilibrium limit. The degree of momentum-space anisotropy ($\xi$) in the range $-1 < \xi < \infty$ is given by 
\begin{equation}
\xi = \frac{1}{2}\frac{\langle \bf k^{2}_{\perp}\rangle}{\langle k^2_z\rangle}-1\, ,
\end{equation}
where $k_z \equiv \bf k \cdot n$ and $\bf k_{\perp}\equiv \bf k-\bf n \, (k\cdot n)$ correspond to the particle momenta along and perpendicular to the direction of anisotropy ($\bf{n}$), respectively. Many prior works have studied heavy quarkonium physics by considering the momentum-space anisotropy inside the QGP~\cite{Strickland:2011aa,Thakur:2013nia,Islam:2020gdv,Islam:2020bnp,Dumitru:2009ni,Romatschke:2003ms, Brambilla:2022ynh, Omar:2021kra}. In this proceeding contribution, we emphasize the efficient consideration of momentum-space anisotropy in a one-dimensional effective theoretical framework and compare the one- and three-dimensional results numerically~\cite{Dong:2021gnb, Dong:2022mbo, Islam:2022qmj}.

\section{Isotropic, an-isotropic, and effective HQ potential}
After considering the perturbative and non-perturbative contributions as shown in our works~\cite{Dong:2022mbo, Islam:2022qmj}, the total complex isotropic HQ potential is given by
\ba
\mathrm{Re}\,V_{\mathrm{Iso}} (r) &=& \alpha m_D \left(\frac{1-e^{-r m_D}}{r m_D}\right)- \alpha m_D - \frac{\sigma}{m_D}\left(2 + r m_D\right)e^{-r m_D} + \frac{2\sigma}{m_D}-\frac{\alpha}{r} \, ,
\ea
\ba
\mathrm{Im}\,V_{\mathrm{Iso}} (r) &=& \alpha \lambda \phi_2\left(r m_D\right) - \alpha \lambda -  \frac{8\sigma\lambda}{m_D^2}\phi_3(r m_D) + \frac{24\sigma\lambda}{m_D^2}\phi_4(r m_D) - \frac{4\sigma\lambda}{m_D^2} \, .
\ea
We also insert a relativistic correction, $-0.8 \sigma /(m_{b/c}^2 r)$, in the potential model while solving the Schr{\"o}dinger equation for charmonia ($m_c=1.3\, {\rm GeV}$) and bottomonia ($m_b=4.7\, {\rm GeV}$ which accounts for subleading corrections to the infinite mass limit.
\paragraph*{}
The real and imaginary parts of the 3D anisotropic potential model used in our previous work~\cite{Dong:2022mbo} are
\begin{equation}
\mathrm{Re}\,V_{\mathrm{Aniso}} (r, \theta, \xi) = \alpha m_D^A \left(\frac{1-e^{-r m_D^R}}{r m_D^R}\right)-\alpha m_D^A - \frac{\sigma}{m_D^A}\left(2 + r m_D^R\right) e^{-r m_D^R}  + \frac{2\sigma}{m_D^A}-\frac{\alpha}{r}\, ,
\end{equation}
	
	\begin{equation}
	\mathrm{Im}\,V_{\mathrm{Aniso}} (r, \theta, \xi) = \alpha \lambda^A \phi_2 \left(r m_D^I\right) - \alpha \lambda^A - \frac{8\sigma \lambda^A}{\left(m_D^A\right)^2}\phi_3 \left(r m_D^I\right) +  \frac{24\sigma \lambda^A}{\left(m_D^A\right)^2}\phi_4 \left(r m_D^I\right) -\frac{4\sigma \lambda^A}{\left(m_D^A\right)^2} \, ,
	\end{equation}
	where,
	\begin{equation}
	m_D^A =m_D \left(1-\frac{\xi}{6}\right)\, ,\quad \lambda^A = \lambda\left(1-\frac{\xi}{6}\right) ,\label{p15}
	\end{equation}
	and
	\begin{equation}
	m^R_D = m_D \bigg[1+\xi\left(0.108\cos 2\theta -0.131\right)\bigg]\, , \quad m^I_D =m_D \bigg[1+\xi\left(0.026\cos 2\theta -0.158\right)\bigg] .\label{p16}
	\end{equation}
\paragraph*{}
We introduce an angle-averaged effective screening mass ${\cal{M}}_{l m}(\lambda,\xi)$ ~\cite{Dong:2021gnb}
\ba\la{effm0}
{\cal{M}}^{R,I}_{l m}(\lambda,\xi)&=&\langle {\rm{Y}}_{l m}(\theta,\phi)| m^{R,I}_D(\lambda,\xi,\theta) | {\rm{Y}}_{l m}(\theta,\phi)\rangle\, ,\nonumber \\
&=&\int_{-1}^{1} d \cos \theta \int_{0}^{2\pi} d \phi {\rm{Y}}_{l m}(\theta,\phi)  m^{R,I}_D(\lambda,\xi,\theta) {\rm{Y}}^*_{l m}(\theta,\phi)\, ,
\ea
where ${\rm{Y}}_{l m}(\theta,\phi)$ are spherical harmonics with azimuthal quantum number $l$ and magnetic quantum number $m$. We obtain the real and imaginary parts of the 1D effective potential model as ~\cite{Dong:2022mbo}
\begin{equation}
\mathrm{Re}\,V_{\mathrm{Eff}} (r,\xi) = \alpha m_D^A \left(\frac{1-e^{-r {\cal{M}}_{lm}^R}}{r {\cal{M}}_{lm}^R}\right)-\alpha m_D^A - \frac{\sigma}{m_D^A}\left(2 + r {\cal{M}}_{lm}^R\right) e^{-r {\cal{M}}_{lm}^R}  + \frac{2\sigma}{m_D^A}-\frac{\alpha}{r} \, ,
\end{equation}

\begin{equation}
\mathrm{Im}\,V_{\mathrm{Eff}} (r,\xi) = \alpha \lambda^A \phi_2 \left(r {\cal{M}}_{lm}^I\right) - \alpha \lambda^A - \frac{8\sigma \lambda^A}{\left(m_D^A\right)^2}\phi_3 \left(r {\cal{M}}_{lm}^I\right) +  \frac{24\sigma \lambda^A}{\left(m_D^A\right)^2}\phi_4 \left(r {\cal{M}}_{lm}^I\right) -\frac{4\sigma \lambda^A}{\left(m_D^A\right)^2} \, ,
\end{equation}
where,
\begin{equation}
K_{lm} = \frac{2l(l+1)-2 m^2 -1}{4l(l+1)-3} \, ,
\end{equation}
and
\begin{equation}
{\cal{M}}_{lm}^R = m_D \bigg[1+\xi\left(0.216 K_{lm} -  0.239\right)\bigg]~,\quad {\cal{M}}_{lm}^I = m_D \bigg[1+\xi\left(0.052 K_{lm} -  0.184\right)\bigg] .
\end{equation}
The $l$ and $m$ values of various quarkonium states are given in Table-1 of \cite{Islam:2022qmj}.
\section{Results}
\begin{figure}[t]
\centering
\includegraphics[width=1\textwidth]{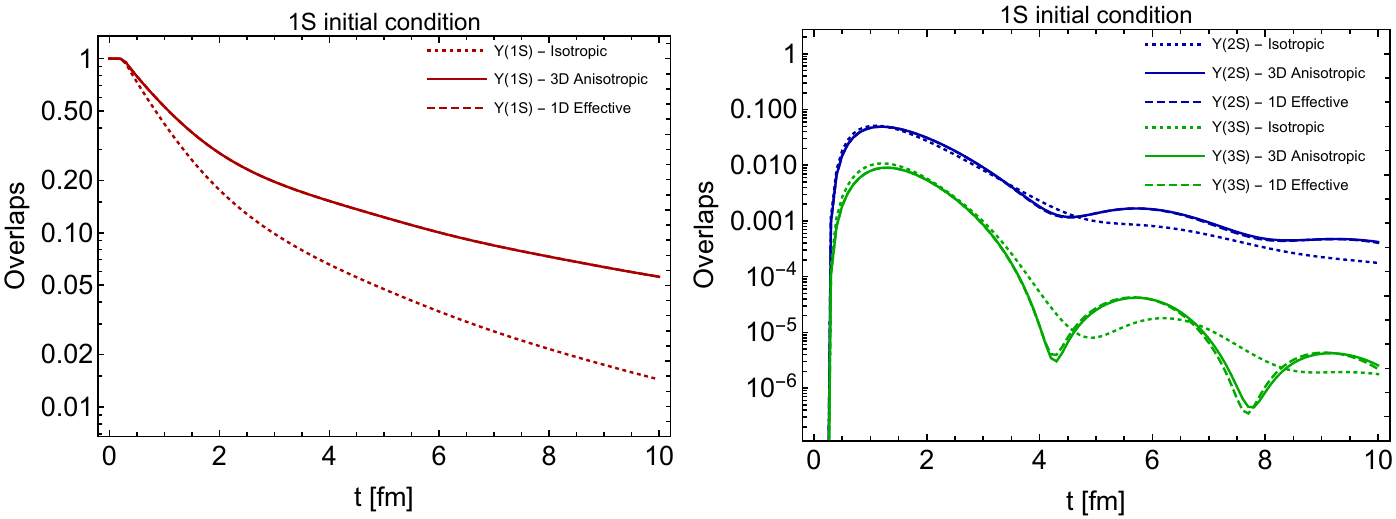}
\caption{The overlaps of $\Upsilon(1S)$, $\Upsilon(2S)$, and $\Upsilon(3S)$ resulting from real-time solution of the Schr\"odinger equation. Here, the wave function has been initialized as pure $\Upsilon(1S)$ eigenstate~\cite{Dong:2022mbo}.}
\label{fig:overlaps-swave}
\end{figure}

To solve the 3D Schr\"odinger equation in real-time, we used a split-step pseudospectral method with temporal step size $\Delta t = 0.001$ fm/c. We compare results obtained with the full 3D anisotropic potential to those obtained with the 1D effective potential. We evolve the wave function from $\tau = 0$ fm/c to $\tau = 0.25$ fm/c in the vacuum ($T=0$).  Starting at $\tau = \tau_0 = 0.25$ fm/c, we consider a fixed anisotropy parameter $\xi = 1$ and boost-invariant Bjorken evolution for the hard scale
\be
\lambda(\tau) = \lambda_0 \left( \frac{\tau_0}{\tau} \right)^{1/3} \, .
\ee
Here we take the initial hard scale to be $\lambda_0 =$ 630 MeV. Further details of the numerical method can be found in ~\cite{Dong:2022mbo}.
We take the box size to be $L =$ 2.56 fm, $m_b = 4.7$ GeV, and use $N=128$ lattice points in each direction for bottomonium states. Fig.~\ref{fig:overlaps-swave} shows the time evolution of overlaps of the $\Upsilon(1S)$, $\Upsilon(2S)$, and $\Upsilon(3S)$ using a pure $\Upsilon(1S)$ eigenstate as the initial condition. The time evolution of the bottomonium p-wave overlaps resulting from initialization with different p-wave polarizations has been shown in Ref.~\cite{Dong:2022mbo}. Results with pure $\Upsilon(2S)$ and $\Upsilon(3S)$ eigenstates and a Gaussian as the initial condition can be found in Ref.~\cite{Dong:2022mbo}. One can also find a similar treatment for charmonium states in Ref.~\cite{Dong:2022mbo}.
\section{Summary}
By introducing an effective screening mass dependent on the quantum numbers $l$ and $m$ of a specific state, we successfully transformed anisotropic heavy-quark potentials into isotropic ones. Our demonstration with the resulting 1D effective potential model revealed the ability to accurately replicate the full 3D outcomes for the energies and binding energies of low-lying heavy-quarkonium bound states. This discovery holds significance as it enables the incorporation of anisotropy effects into one-dimensional real-time Schrödinger equations, forming the foundation for phenomenological calculations related to bottomonium suppression in open quantum systems approaches.

\vspace{6pt} 





\funding{This work was funded by the U.S. Department of Energy, Office of Science, Office of Nuclear Physics Award No.~DE-SC0013470..}

\acknowledgments{This contribution is based on two papers listed in Refs.~\cite{Dong:2021gnb, Dong:2022mbo} which were published in collaboration with L. Dong, Y. Guo, A. Rothkopf, and M. Strickland. The speaker (A.I.) would like to thank the organizers of the Hot Quarks 2022 conference for the opportunity to present this talk.}

\conflictsofinterest{The authors declare no conflict of interest.} 





\appendixtitles{no} 

\begin{adjustwidth}{-\extralength}{0cm}




\reftitle{References}
\externalbibliography{yes}
\bibliography{main.bib}

%


\PublishersNote{}
\end{adjustwidth}
\end{document}